\documentclass{aa} 

\usepackage{graphicx}
\usepackage{txfonts}
\usepackage{CJK}
\usepackage{enumitem}
\usepackage{amsmath}
\usepackage{amssymb}
\usepackage{natbib}
\bibpunct{(}{)}{;}{a}{}{,}
\usepackage{url}
\usepackage{hyperref}
\hypersetup{
    colorlinks=true,
    linkcolor=blue,
    filecolor=blue,      
    urlcolor=blue,
    citecolor=blue
}
\newcommand   {\Teff}    {T_{\rm eff}}
\newcommand   {\logg}    {{\rm log}\,g}
\newcommand   {\meta}    {{\rm [M/H]}}

\newcommand   {\Vrad}    {V_{\rm rad}}
\newcommand   {\kms}     {\rm km\,s^{-1}}

\newcommand   {\micron}  {\rm \mu m}

\makeatletter
\renewcommand*\aa@pageof{, page \thepage{} of \pageref*{LastPage}}
\makeatother
\usepackage{color}

\begin{document}

\begin{CJK*}{UTF8}{gbsn}

\title{Solid confirmation of the broad DIB around 864.8 nm using stacked {\it Gaia}--RVS spectra}

\author{H. Zhao (赵赫)\inst{1,2}
        \and
        M. Schultheis\inst{1}
        \and
        T. Zwitter\inst{3}
        \and
        C.A.L. Bailer-Jones\inst{4} 
        \and
        P. Panuzzo\inst{5} 
        \and
        P. Sartoretti\inst{5} 
        \and 
        G.M. Seabroke\inst{6} 
        \and 
        A. Recio-Blanco\inst{1}
        \and
        P. de Laverny\inst{1}
        \and
        G. Kordopatis\inst{1}
        \and
        O.L. Creevey\inst{1}
        \and
        T.E. Dharmawardena \inst{4} 
        \and 
        Y. Fr\'emat\inst{7}
        \and
        R. Sordo \inst{8}
        \and
        R. Drimmel \inst{9} 
        \and
        D.J. Marshall \inst{10} 
        \and 
        P.A. Palicio\inst{1}
        \and
        G. Contursi\inst{1}
        \and
        M.A. \'Alvarez\inst{11} 
        \and
        S. Baker\inst{6}
        \and
        K. Benson\inst{6}
        \and
        M. Cropper\inst{6}
        \and
        C. Dolding\inst{6}
        \and
        H.E. Huckle\inst{6}
        \and
        M. Smith\inst{6}
        \and
        O. Marchal\inst{12} 
        \and
        C. Ordenovic\inst{1}
        \and
        F. Pailler\inst{13}
        \and
        I. Slezak\inst{1}
        }

\institute{University C\^ote d'Azur, Observatory of the C\^ote d'Azur, CNRS, Lagrange
          Laboratory, Observatory Bd, CS 34229, \\
          06304 Nice cedex 4, France \\
          \email{he.zhao@oca.eu, mathias.schultheis@oca.eu}
          \and
          Purple Mountain Observatory, Chinese Academy of Sciences, Nanjing 210023, PR China
          \and
          Faculty of Mathematics and Physics, University of Ljubljana, Jadranska 19, 1000 Ljubljana, Slovenia
          \and
          Max Planck Institute for Astronomy, K\"{o}nigstuhl 17, 69117 Heidelberg, Germany
          \and
          GEPI, Observatoire de Paris, Universit\'{e} PSL, CNRS, 5 Place Jules Janssen, 92190 Meudon, France
          \and
          Mullard space science laboratory, University College London, Holmbury St Mary, Dorking, Surrey, RH5  6NT, United Kingdom
          \and
           Royal Observatory of Belgium, 3 avenue circulaire, 1180 Brussels, Belgium
           \and
          INAF - Osservatorio astronomico di Padova, Vicolo Osservatorio 5, 35122 Padova, Italy
          \and
          INAF - Osservatorio Astrofisico di Torino, via Osservatorio 20, 10025 Pino Torinese (TO), Italy
          \and
          IRAP, Universit\'{e} de Toulouse, CNRS, UPS, CNES, 9 Av. colonel Roche, BP 44346, 31028 Toulouse Cedex 4, France
          \and
          CIGUS CITIC - Department of Computer Science and Information Technologies, University of A Coru\~{n}a, Campus de Elvi\~{n}a s/n, A Coru\~{n}a, 15071, Spain
          \and
          Universit\'{e} de Strasbourg, CNRS, Observatoire astronomique de Strasbourg, UMR 7550, 11 rue de l'Universit\'{e}, 67000 Strasbourg, France
          \and
          CNES Centre Spatial de Toulouse, 18 avenue Edouard Belin, 31401 Toulouse Cedex 9, France
      }

\date{Received ??; accepted ??}


 
\abstract
{Studies of the correlation between different diffuse interstellar bands (DIBs) are important for exploring their origins. However,
the {\it Gaia}--RVS spectral window between 846 and 870\,nm contains few DIBs, the strong DIB at 862\,nm being the only 
convincingly confirmed one.}
{Here we attempt to confirm the existence of a broad DIB around 864.8\,nm  and estimate its characteristics using the stacked {\it 
Gaia}--RVS spectra of a large number of stars. We study the correlations between the two DIBs at 862\,nm ($\lambda$862) and 864.8\,nm
($\lambda$864.8), as well as the interstellar extinction.}
{We obtained spectra of the interstellar medium (ISM) absorption by subtracting the stellar components using templates constructed 
from real spectra at high Galactic latitudes with low extinctions. We then stacked the ISM spectra in Galactic coordinates ($\ell,\,b$) 
---pixelized by the HEALPix scheme--- to measure the DIBs. The stacked spectrum is modeled by the profiles of the two DIBs, 
Gaussian for $\lambda$862 and Lorentzian for $\lambda$864.8, and a linear continuum. We report the fitted central depth (CD), 
central wavelength, equivalent width (EW), and their uncertainties for the two DIBs.}
{We obtain 8458 stacked spectra in total, of which 1103 (13\%) have reliable fitting results after applying numerous conservative 
filters. This work is the first of its kind to fit and measure $\lambda$862 and $\lambda$864.8 simultaneously in cool-star spectra. 
Based on these measurements, we find that the EWs and CDs of $\lambda$862 and $\lambda$864.8 are well correlated with each other, 
with Pearson coefficients ($r_p$) of 0.78 and 0.87, respectively. The full width at half maximum (FWHM) of $\lambda$864.8 is estimated 
as $1.62\pm 0.33$\,nm which compares to $0.55\pm 0.06$\,nm for $\lambda$862. We also measure the vacuum rest-frame wavelength of 
$\lambda$864.8 to be $\lambda_0\,{=}\,864.53\pm0.14$\,nm, smaller than previous estimates.} 
{We find solid confirmation of the existence of the DIB around 864.8\,nm based on an exploration of its correlation with $\lambda$862 
and estimation of its FWHM. The DIB $\lambda$864.8 is very broad and shallow. That at $\lambda$862 correlates better with $\rm E(BP-RP)$ 
than $\lambda$864.8. The profiles of the two DIBs could strongly overlap with each other, which contributes to the skew of the 
$\lambda$862 profile.}

\maketitle
%

\section{Introduction}

Diffuse interstellar bands (DIBs) are a set of absorption features with profiles that are much broader than those of the 
interstellar atomic lines (e.g., \ion{Na}{i} lines). To date, over 600 DIBs have been identified at optical and near-infrared 
wavelengths (0.4--2.4\,$\micron$; \citealt{Galazutdinov2017b}, \citealt{Fan2019}). However, our knowledge about their origins is 
still very limited. So far, only $C^{+}_{60}$ has been confirmed as the carrier for five DIBs between 930 and 965\,nm (see 
\citealt{Linnartz2020} for a review). Gas-phase carbon-bearing molecules are thought to be the most probable DIB carriers 
\citep[e.g.,][]{Snow2014iaus}.

Despite the unknown nature of DIBs, absorption intensity maps have been built for several such features detected in large spectroscopic 
surveys \citep[e.g.,][]{Kos2014,Zasowski2015c,Baron2015a,Lan2015,Puspitarini2015} in order to explore the interstellar medium (ISM) 
and Galactic structure. Among them, the latest map was built for the DIB at 862\,nm (DIB\,$\lambda862$) using data from the {\it 
Gaia} Radial Velocity Spectrometer \citep[RVS;][]{Seabroke2022} in Data Release 3, which has the largest sky coverage to date (see 
\citealt{Schultheis2022} for more details). Their wavelengths and line widths are expressed in nanometers following the unit used in 
the {\it Gaia} analysis, while their strength is still expressed in \r{A}ngstr\"oms due to their small line depth ($\lesssim$6\%). 
The mentioned wavelengths in this paper are in the vacuum. 
 
Studies of the correlation between different DIBs are important for identifying their origins \citep[e.g.,][]{Friedman2011,Elyajouri2017b}. 
However, such studies are difficult for $\lambda$862 because the RVS spectral window (846--870\,nm) has very few DIBs. In the review 
of possible DIBs in this region by \citet{Munari2008}, the most promising one was thought to be located around 864.8\,nm ($\lambda$864.8) 
with a very shallow profile. $\lambda$864.8 was first reported by \citet{Sanner1978} following positive support from \citet{HL1991}, 
\citet{JD1994}, and \citet{Wallerstein2007}. This DIB was clearly seen in spectra of early-type stars, but its analysis is complex 
because of the wing of the hydrogen Paschen 13 line and other stellar components. On the other hand, this band was not reported by 
\citet{Fan2019}. \citet{Krelowski2019b} argued that this feature was of stellar origin (\ion{He}{i} lines) as it did not correlate 
with dust extinction. However, \citet{Baron2015b} confirmed weak correlations between dust extinction and many DIBs, which weakens 
this argument. Furthermore, a position correlation with dust extinction could be seen around 865\,nm in Fig. 2 in \citet{Baron2015b},
strengthening the DIB hypothesis for this feature.

In this work, we use the unique set of {\it Gaia}--RVS spectra of cool stars observed over the whole sky which has just been published 
as part of the third {\it Gaia} data release \citep[DR3;][]{Gaia-mission,Vallenari2022}. We aim to confirm $\lambda$864.8 as a broad 
DIB rather than a stellar component by the use of stacked {\it Gaia}--RVS spectra to detect it and then analysis of its correlations 
with dust extinction and DIB\,$\lambda$862. 

\section{Data} \label{sect:data}

The {\it Gaia}--RVS spectra have a wavelength coverage of [$846-870$]\,nm and a medium resolution of $R\,{\sim}\,11\,500$ \citep{Cropper2018}.
The stellar parameters and chemical abundances, including effective temperature ($\Teff$), surface gravity ($\logg$), mean metallicity 
($\meta$), and individual abundances of up to 13 chemical species, are derived by the General Stellar Parametrizer from spectroscopy 
({\it GSP-Spec}) module of the Astrophysical parameters inference system \citep[Apsis,][]{Creevey2022,Recio-Blanco2022}. We refer to 
\citet{Recio-Blanco2022} for detailed descriptions of the parametrization and a validation of the measurement of $\lambda$862 in 
individual RVS spectra.

About one million {\it Gaia}--RVS spectra have been published in {\it Gaia} DR3. We use a sample of 648\,944 spectra that meet the 
following criteria (given as an ADQL query example in Appendix \ref{queries}):

\begin{enumerate}
    \item Stellar parameters ($\Teff$, $\logg$, $\meta$) and radial velocity ($\Vrad$) are not NaN values.
    \item $3500\,{\rm K}\,{\leqslant}\,\Teff\,{\leqslant}\,7500$\,K and $\logg\,{<}\,6$\,dex.
    \item Stellar distances derived from parallax measurements ($1/\varpi$) are within 6\,kpc.
    \item The signal-to-noise ratios (S/N) of the observed spectra are greater than 20.
    \item The uncertainty of $\Vrad$ is smaller than 5\,$\kms$.
\end{enumerate}

This work processes stars with $\Teff\,{\leqslant}\,7500$\,K, defined as ``cool stars''. The whole sample is separated into two 
parts: 75\,941 ``reference spectra'' at high latitudes ($|b|\,{\geqslant}\,30^{\circ}$) and with low extinctions 
($\rm E(BP-RP)\,{<}\,0.02$\,mag) are used to construct the stellar templates for the cool stars; the other 573\,003 are the 
``target spectra'' in which we attempt to fit and measure the two DIBs. The reddening $\rm E(BP-RP)$ used in this work was derived 
by the {\it GSP-Phot} module using their BP$/$RP spectra \citep{Andrae2022}. Our sample uses about one-tenth of the total RVS spectra 
processed by {\it GSP-Spec} \citep{Recio-Blanco2022}. \citet{Kos2013} and \citet{Lan2015} made use of a similar number of spectra in 
their studies. \citet{Baron2015b,Baron2015a} used over 1.5 million  extragalactic spectra. The large samples enable us to measure 
very weak DIBs by binning the spectra in large numbers.

\begin{figure}[!ht]
  \centering
  \includegraphics[width=8cm]{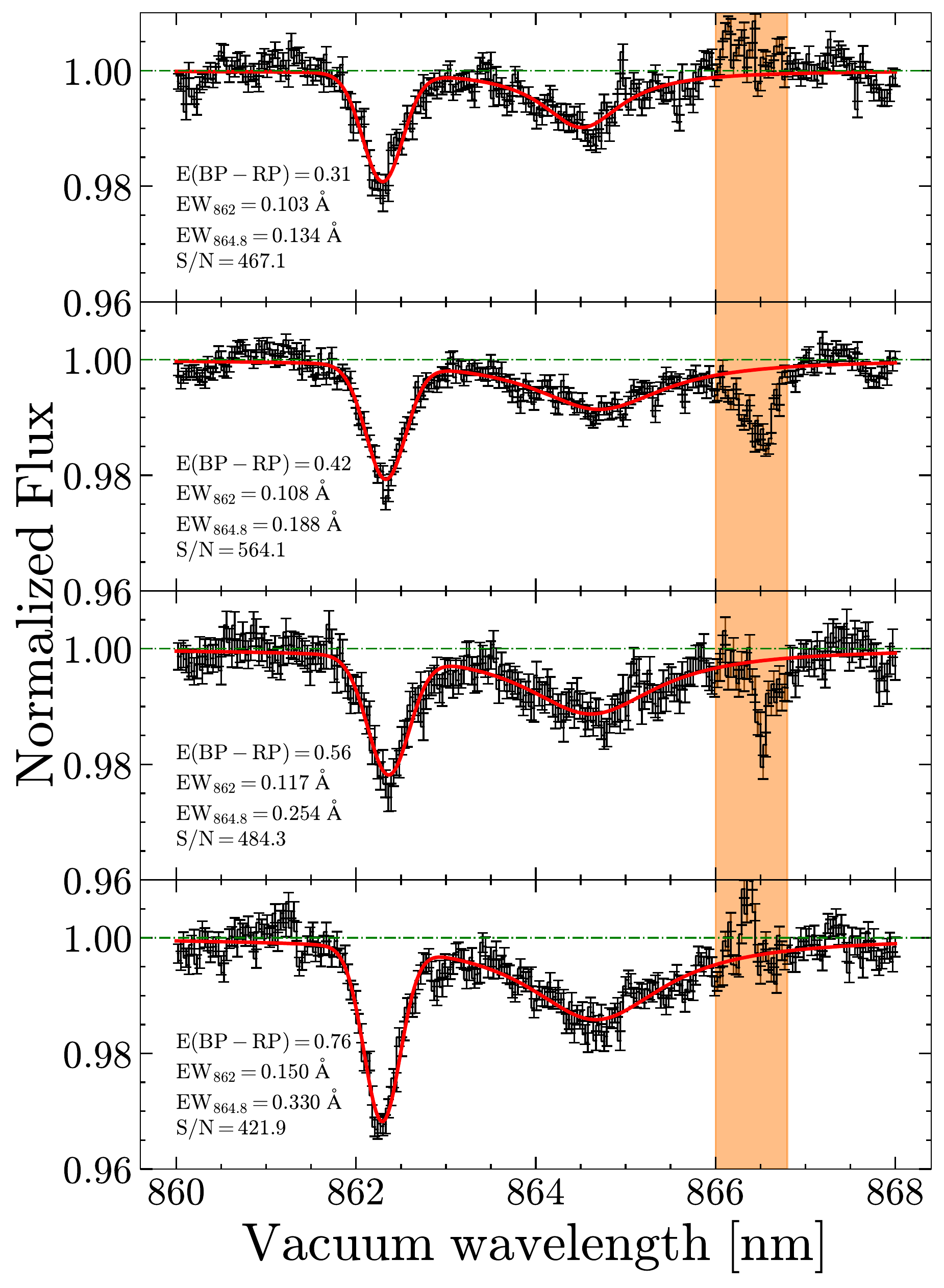}
  \caption{Four examples of the fits to DIBs $\lambda862$ and $\lambda864.8$ in stacked {\it Gaia}--RVS spectra. The black lines are 
  the ISM spectra normalized by the fitted linear continuum. The error bars indicate the flux uncertainties at each pixel. The red 
  lines are the fitted curves that are normalized by the continuum as well. Orange marks the \ion{Ca}{ii} line residuals. The median 
  $\rm E(BP-RP)$ and EWs of the two DIBs in each HEALPix are also indicated.}
  \label{fig:fit-eg}
\end{figure}

\section{Method} \label{sect:fit-dib}

This work is the first of its kind to fit and measure $\lambda$862 and $\lambda$864.8 simultaneously in cool-star spectra. To do so, 
first we derive the ISM spectrum for each target by subtracting their stellar components using the reference spectra. We then stack 
the ISM spectra of the targets according to their Galactic coordinates ($\ell,\,b$), per HEALPix. Finally we fit and measure the two 
DIBs, $\lambda862$ and $\lambda864.8$, in the stacked spectra.

The first step follows the main principles of \citet{Kos2013}. For a given target, we find a set of reference spectra with similar 
parameters to the {\it GSP-spec} values of the target that have $\Teff\,{\pm}\,20\%$, $\logg\,{\pm}\,0.6$\,dex, and $\meta\,{\pm}\,0.4$\,dex. 
These ranges are arbitrary, but chosen to be larger than the parameter uncertainties \citep{Kos2013}. The constraints on the stellar 
parameters are not necessary but they can speed up the procedure. The similarity between the target spectrum and each reference 
spectrum is then calculated as the average absolute difference of their flux at all pixels, except a masked region of 860--868\,nm 
where the two DIBs are located. When measuring the difference, the central regions of the \ion{Ca}{ii} triplet are down-weighted to 
30\% because \ion{Ca}{ii} lines dominate the whole {\it Gaia}--RVS spectra of cool stars whereas the similarity between \ion{Fe}{i} 
lines (close to DIBs) is more important than between the \ion{Ca}{ii} lines for constructing the templates (see examples in 
\citealt{Contursi2021}). In practice, the two \ion{Ca}{ii} regions of concern are defined as 849.43--851.03\,nm and 853.73--855.73\,nm 
and are down-weighted. The \ion{Ca}{ii} line near 866.5\,nm is within the masked DIB region. Reference spectra with average differences 
of greater than the inverse of the square of the target S/N are discarded. For the rest, up to 500 (and at least 10) best-matching 
reference spectra are averaged and weighted by their S/N in order to build a stellar template. The target spectrum divided by the 
template then gives the ISM spectrum for this target. An illustration of this method can be found in Fig. 2 in \citet{Kos2013}. Examples 
of RVS spectra and successful measurements of the DIB\,$\lambda862$ in individual RVS spectrum can be found in \citet{Recio-Blanco2022}.

In the second step, the ISM spectra are stacked according to their Galactic coordinates ($\ell,\,b$) to get a sufficiently high S/N  
for a reliable measurement of $\lambda$864.8. The pixelation of the sky is done by the HEALPix\footnote{\url{https://healpix.sourceforge.io}} 
\citep{Gorski2005} scheme. We choose level 5 ($N_{\rm side}\,{=}\,32$), corresponding to 12\,288 pixels and a spatial resolution about 
$1.8^{\circ}$. We note that 8458 HEALPix pixels (69\% of the total) contain at least ten targets that have ISM spectra ($N_{\rm tar}\,{>}\,10$).
In each of these pixels, we stack all of the ISM spectra (54 on average with a maximum of 362) to generate one stacked spectrum. 
We stack the ISM spectra in each pixel by taking the median value of the fluxes in order to reduce the influence of outliers. The flux 
uncertainty of the stacked spectrum is the mean of the individual flux uncertainties divided by $\sqrt{N_{\rm tar}}$, which is the 
standard error in the mean. The S/N of the stacked spectra is calculated between 860.2 and 861.2\,nm as $\rm mean(flux)/std(flux)$. 
We have 8458 stacked spectra in total, one for each HEALPix pixel.

In the third and final step, a Markov chain Monte Carlo (MCMC) procedure \citep{Foreman-Mackey13} is used to fit each stacked 
spectrum between 860 and 868\,nm (the DIB region) with a Gaussian function for the profile of $\lambda$862, a Lorentzian function 
(better than Gaussian when considering the goodness of fit to the line wings) for the profile of $\lambda$864.8, and a linear function 
for the continuum placement, masking 866--866.8\,nm for the \ion{Ca}{ii} line residuals. The best estimates and statistical uncertainties 
are taken in terms of the 50th, 16th, and 84th percentiles of the posterior distribution. We note that the \ion{Ca}{ii} region slightly 
changes after the shifting to the heliocentric frame, considering that the radial velocities of most stars lie mainly within $\pm$50\,$\kms$, 
corresponding to $|\Delta \lambda|\,{\simeq}\,0.14$\,nm at 866.5\,nm. Furthermore, the wings of the \ion{Ca}{ii} lines are better 
modeled than their central parts \citep{Creevey2022,Recio-Blanco2022}. Thus, our masked region can properly prevent contamination by
the \ion{Ca}{ii} residuals and consequently the $\lambda$864.8 profile can be fitted well. Figure \ref{fig:fit-eg} shows four fitting 
examples sorted by the median $\rm E(BP-RP)$ in each HEALPix pixel. The fitting results for all stacked spectra are available online.

Besides the fitted central depth (CD) and central wavelength ($\lambda_C$) of the DIB profiles, the full width at half maximum (FWHM)
and equivalent width (EW) are also calculated. The EW errors are estimated by the span of the profile (3$\times$FWHM), the pixel 
resolution ($\delta \lambda\,{=}\,0.03$\,nm$/$pixel), and the noise level of the line center ($R_C\,{=}\,{\rm std(data-model)}$), 
as ${\rm \Delta EW}\,{=}\,\sqrt{6\,{\rm FWHM}\,\delta \lambda} \times {\it R_C}$, similar to the formulas given by \citet{Vos2011} 
and \citet{VE2006}. 

\begin{figure}[!ht]
  \centering
  \includegraphics[width=8cm]{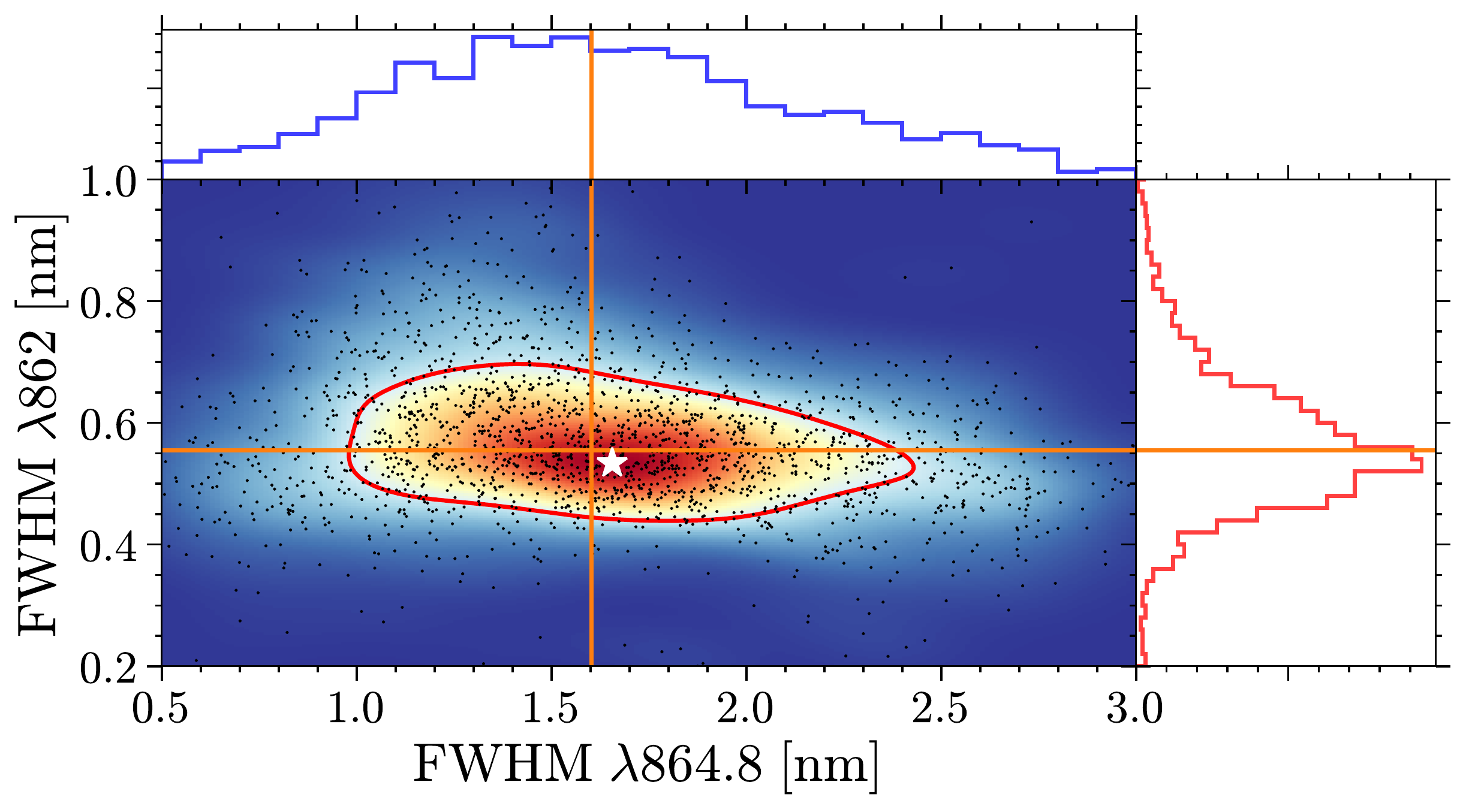}
  \caption{Distributions of the FWHM of $\lambda862$ (red histogram) and $\lambda864.8$ (blue histogram), as well as their joint 
  distribution (middle colored map), measured in 1962 stacked spectra after applying the general filtering. The colors represent 
  the densities of the joint FWHM distribution, estimated by a Gaussian KDE. The white star indicates the peak density. The red 
  line in the central panel indicates a contour with a probability density of 1.2, about one-third of the peak density. The orange 
  lines indicate the median FWHM of $\lambda862$ and $\lambda864.8$, respectively.}
  \label{fig:width}
\end{figure}

\section{Results} \label{sect:results}

The aforementioned fit of the DIB was done for each of the 8458 stacked spectra (one per HEALPix pixel). To get reliable results, 
we only retain the stacked spectra that meet the following criteria, which leaves us with 1962 spectra:

\begin{itemize}
    \item S/N of the stacked ISM spectrum is greater than 100.
    \item $\rm CD_{862} > 3{\it R}_C$ and $\rm CD_{864.8} > {\it R}_C$. 
    \item The FWHM of $\lambda862$ and $\lambda864.8$ are greater than 0.2\,nm and 0.5\,nm, respectively.
\end{itemize}

Figure \ref{fig:width} shows the distributions of the FWHM of $\lambda862$ and $\lambda864.8$, as well as their joint distribution, 
for the 1962 stacked spectra. The FWHM of\,$\lambda864.8$ has a much wider distribution than that of $\lambda$862, and both slightly 
deviate from a Gaussian. We apply a Gaussian kernel density estimation (KDE) to their joint distribution with a bandwidth of 0.2826\,nm 
(automatically determined by the Python package {\it scipy}). The median FWHM is to the upper left of the peak density estimated by 
the KDE (white star in Fig. \ref{fig:width}), which is caused by the fits to some very shallow profiles that result in broader 
$\lambda$862 and narrower $\lambda$864.8. The outliers in low-density regions are due to the noise in stacked spectra.

Therefore, we discard the points that lie in the region with a density of less than 1.2 (red line in Fig. \ref{fig:width}), which 
is about one-third of the peak density. After this final filtering, our sample comprises 1103 reliable measurements of the two DIBs, 
about 13\% of the total stacked spectra.

\begin{figure}[!ht]
  \centering
  \includegraphics[width=8cm]{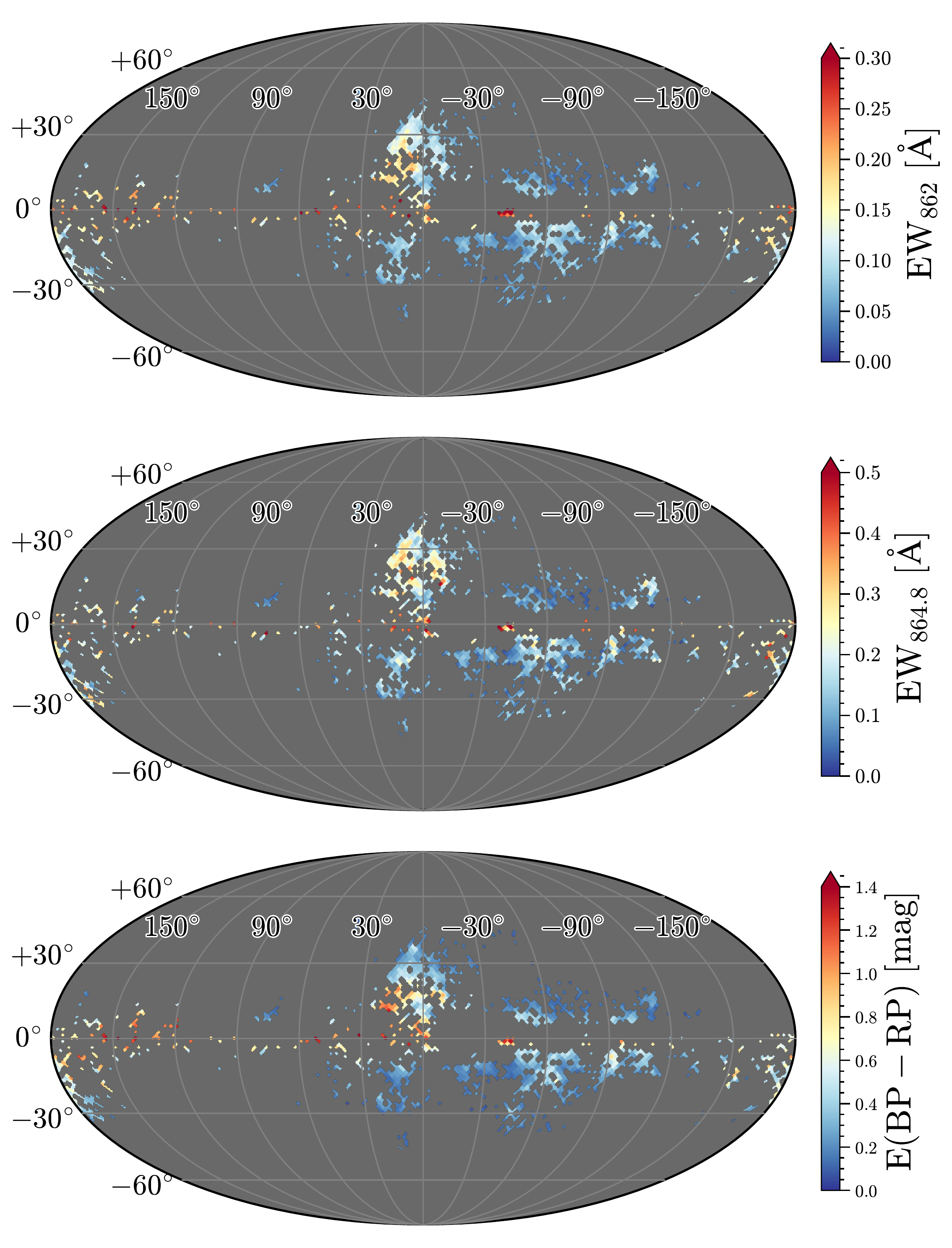}
  \caption{Galactic distributions of the 1103 reliable measurements of the strength of $\lambda862$ (top), $\lambda864.8$ (middle), 
  and $\rm E(BP-RP)$ (bottom), in Mollweide projection, with the Galactic center in the middle and Galactic longitude increasing 
  to the left.}
  \label{fig:2D-map}
\end{figure}

The two-dimensional intensity map of $\lambda862$ and $\lambda864.8$ as well as $\rm E(BP-RP)$ are shown in Galactic coordinates 
in Fig.~\ref{fig:2D-map}. Limited by our sample, the spatial distributions of $\rm EW_{862}$ and $\rm EW_{864.8}$ cannot be well 
described. However, it is clear that large EWs for both $\lambda862$ and $\lambda864.8$ concentrate in the Galactic plane. 
Furthermore, decreasing $\rm EW_{862}$ and $\rm EW_{864.8}$ with latitude (up to $b\,{=}\,{\pm}30^{\circ}$) can be found near 
Galactic center (GC, $|\ell|\,{<}\,30^{\circ}$), which are similar to each other and that of $\rm E(BP-RP)$.

\begin{figure}[!ht]
  \centering
  \includegraphics[width=8cm]{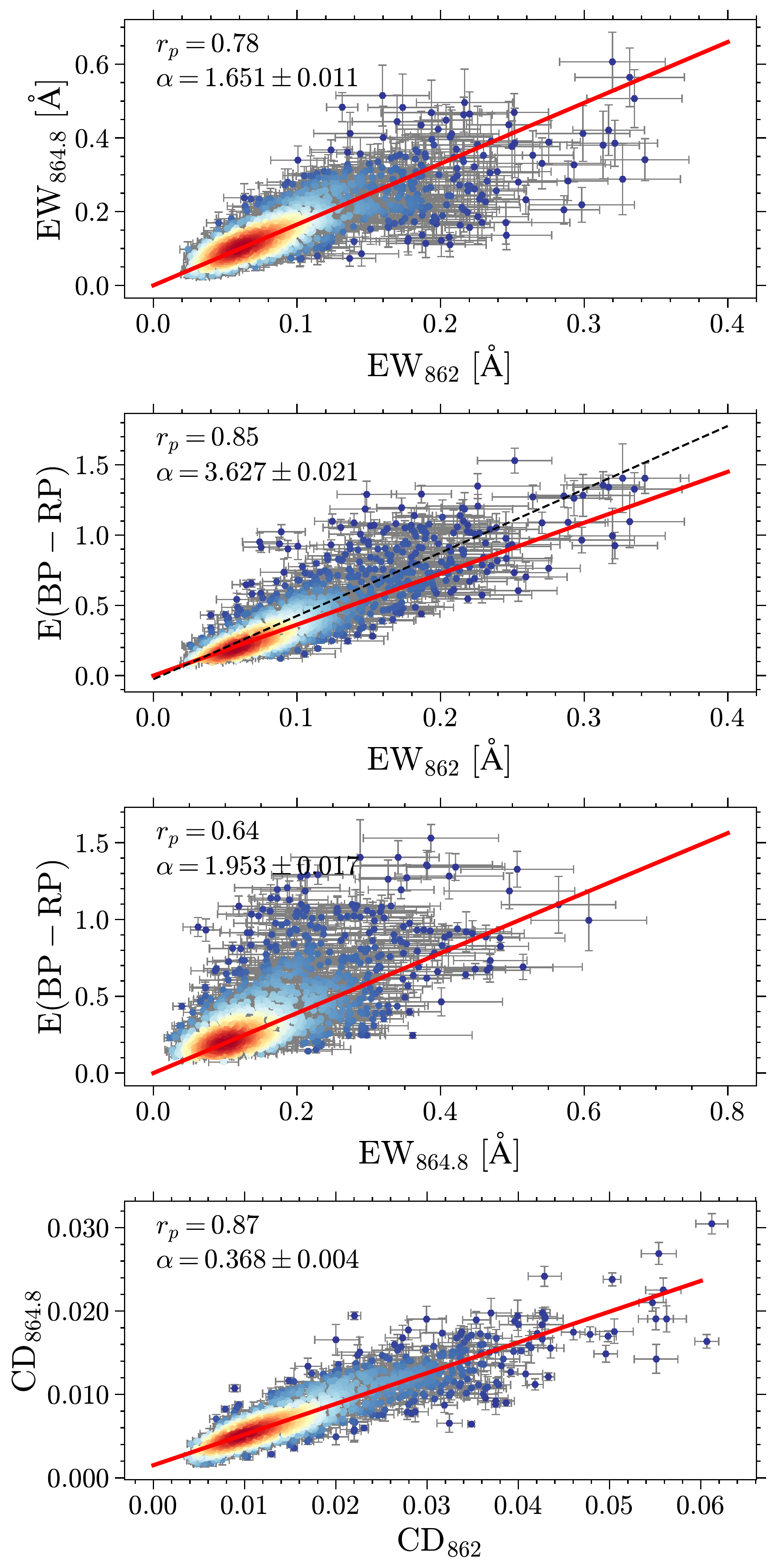}
  \caption{Intensity correlations between $\lambda862$, $\lambda864.8$, and $\rm E(BP-RP)$. Different correlations are shown in 
  each panel with the Pearson coefficient ($r_p$) and the slope of a 2$\sigma$-clipped linear fit ($\alpha$) with zero intercept 
  shown in red. Data points with error bars are colored by their number density estimated by a Gaussian KDE. The black dashed line 
  shown in the second panel from the top is the linear relation derived in \citet{Schultheis2022}.}
  \label{fig:intensity}
\end{figure}

Intensity correlations between $\lambda862$, $\lambda864.8$, and $\rm E(BP-RP)$ are shown in Fig. \ref{fig:intensity}. For $\rm 
E(BP-RP)$, we use $\rm std(E(BP-RP))/\sqrt{N}$ to represent its uncertainty in each HEALPix pixel. $\rm EW_{862}$ and $\rm EW_{864.8}$ 
are well correlated with each other with a Pearson coefficient of $r_p\,{=}\,0.78$. A systematic deviation can be seen when $\rm 
EW_{862}\,{\gtrsim}\,0.2$\,{\AA}. The cause of this is unclear: either it has a physical origin or is due to the FWHM dispersion 
seen in Fig. \ref{fig:width}. To avoid this effect, we apply simple linear fits with 2$\sigma$ clipping and zero intercept for EWs 
and dust reddening, resulting in the following relations: 

\begin{equation} \label{eq:corre}
\begin{split}
    \rm EW_{864.8} & = 1.651(\pm 0.011)  \times  {\rm EW_{862}  }   \\
    \rm E(BP-RP)   & = 3.627(\pm 0.021)  \times  {\rm EW_{862}  }   \\
    \rm E(BP-RP)   & = 1.953(\pm 0.017)  \times  {\rm EW_{864.8}}   \\
    \rm CD_{864.8} & = 0.368(\pm 0.004)  \times  {\rm CD_{862}} + 0.002(\pm 0.0001)
\end{split}
.\end{equation}

We also apply the linear fits without any clipping. For the two DIBs, the coefficient from the no-clipping fit becomes 6\% lower 
($1.548\,{\pm}\,0.015$) for their EWs and 5\% lower ($0.350\,{\pm}\,0.006$) for their CDs, compared to the 2$\sigma$-clipped results. 
The relative strength between DIB and dust is 13\% above for $\lambda$862 ($\rm E(BP-RP)/EW_{862}\,{=}\,4.096\,{\pm}\,0.038$) and 
21\% above for $\lambda$864.8 ($\rm E(BP-RP)/EW_{864.8}\,{=}\,2.367\,{\pm}\,0.034$), respectively, when not clipping. This is not 
surprising, as significant scatter about the fit can be found around $\rm E(BP-RP)\,{\sim}\,1$\,mag and $\rm EW_{862}\,{\sim}\,0.1$\,{\AA}. 
Similar scatter exists for $\lambda$864.8 as well, but this latter is more significant. This scattering could be a result of the 
overestimation of $\rm E(BP-RP)$ as seen in Fig. 7 in \citet{Schultheis2022}. We prefer the 2$\sigma$-clipped results because the 
scattering is not only caused by the measurement uncertainties but may indicate a spatial disconnection between their carriers and 
dust grains. Thus, the fitted coefficients with a strong filtering would represent their mean relative strength in the regions where 
these ISM materials are well mixed.

Compared to EWs, the CDs of the two DIBs are better correlated with each other with $r_p\,{=}\,0.87$ and no systematic deviations.
The depth of $\lambda$864.8 is about 37\% of that of $\lambda$862, while its EW is over 1.6 times $\rm EW_{862}$. We note that $\rm 
EW_{862}$ shows good correlation with $\rm E(BP-RP)$ with $r_p\,{=}\,0.85$ as expected, but their relative strength is about 20\% below 
that derived from the DIB sample in {\it Gaia}--DR3 \citep[$4.507 \pm 0.137$,][]{Schultheis2022}. The reason could be that different 
templates ---from {\it GSP-Spec} and from reference spectra--- would introduce differences in EW measurements (see e.g., 
\citealt{Elyajouri2019}). It should also be noted that $\rm E(BP-RP)/EW_{862}$ fitted in this work with no clipping is closer to the 
result in \citet[][9\% difference]{Schultheis2022} where the authors did not make any filtering either. Also, $\lambda$864.8 does not 
correlate as well as $\lambda$862 with $\rm E(BP-RP),$ showing a much smaller $r_p\,{=}\,0.64$ and a larger dispersion. 

In a similar manner to \citet{Munari2008} and \citet{hz2021b}, in order to estimate the rest-frame wavelength ($\lambda_0$) of 
$\lambda$864.8, we assume that the median radial velocity in the Local Standard of Rest of the DIB carrier is zero toward the GC. 
There are 43 cases with $|\ell|\,{\leqslant}\,10^{\circ}$, $|b|\,{\leqslant}\,10^{\circ}$, and small uncertainties in $\lambda_C$ 
($<$0.1\,nm for $\lambda$862 and $<$0.15\,nm for $\lambda$864.8). For $\lambda$862, we derive a $\lambda_0$ of $862.319\,{\pm}\,0.018$\,nm, 
which is highly consistent with the result of $862.323\,{\pm}\,0.0019$\,nm in \citet{Schultheis2022}. For $\lambda$864.8, we derive 
$\lambda_0\,{=}\,864.53\,{\pm}\,0.14$\,nm in the vacuum, corresponding to 864.29\,nm for air wavelength. This result is smaller than 
the commonly suggested air wavelength measured in the spectra of early-type stars, such as 865.0\,nm \citep{Sanner1978}, 864.9\,nm 
\citep{HL1991}, and 864.83\,nm \citep{JD1994}. 

\section{Discussion} \label{sect:discuss}

Until now, the FWHM of the $\lambda$864.8 feature has not been well determined. \citet{Sanner1978} noted that this DIB profile 
extends from 863 to 866\,nm and its strength did not correlate with spectral type, but they did not measure its FWHM. We found two 
measurements of FWHM\,$\lambda$864.8 in the literature that are dramatically different from each other: 1.4\,nm by \citet{HL1991} 
and 0.42\,nm by \citet{JD1994}. The difficulty is that this band is shallow and superposed across several blended stellar lines, 
such as the \ion{He}{i} line at 864.83\,nm in early B-type supergiants \citep{HL1991}. This may be the reason why a stellar origin 
has been claimed in the literature \citep{Krelowski2019b}. In this work, we made use of the spectra of cool stars, and subtracted 
the stellar lines (mainly \ion{Fe}{i} and \ion{Ca}{ii}) using reference spectra. This makes our measurements more accurate than those 
previous works based on hot-star spectra. We find $\lambda$864.8 to be a very broad and shallow DIB with a FWHM of $1.62\,{\pm}\,0.33$\,nm, 
which is not significantly less than the strongest DIB at 442.8\,nm (e.g., 1.7\,nm; \citealt{Galazutdinov2020}). This provides strong 
evidence for the DIB interpretation of this line because no known stellar components or Doppler broadening (a velocity dispersion 
over 600\,$\kms$) can explain such a large width. On the other hand, the derived FWHM of $\lambda$862 in this work is $0.55\,{\pm}\,0.06$\,nm, 
which is slightly larger than previously reported values such as 0.43\,nm \citep{HL1991}, 0.47\,nm \citep{Maiz-Apellaniz2015a}, and 
0.37\,nm \citep{Fan2019}. The increase in FWHM could be explained by a Doppler broadening caused by our stacking strategy, with a 
velocity dispersion about 30--50\,$\kms$. This value is consistent with the average magnitude of the $\Vrad$ dispersion in each pixel 
(about 36\,$\kms$) which can be treated as an upper limit to the velocity dispersion of the DIB clouds, because closer objects should 
have a smaller radial velocity than the background stars.

Although the DIB profile could be broadened due to the superposition effect, our stacking has only a very weak influence on EW 
measurements, because DIBs are weak spectral features and we are therefore in a linear regime where the EW produced by multiple DIB 
clouds can be simply added up \citep{MZ1997}. The linear correlation between $\rm EW_{862}$ and $\rm EW_{864.8}$ provides good evidence 
for a DIB interpretation of this line, in spite of the superposition effect, because the stellar components or their residuals cannot 
have such a behavior with interstellar features. Another important point is that the profiles of the two DIBs could be overlapped 
with each other. Therefore, it is better to measure the two DIBs simultaneously; although this requires a very high S/N.

Limited by our sample and the conservative approach to filtering it, further analysis of the two DIBs ---such as the kinematics of 
their carriers and the nature of the Lorentzian profile of $\lambda$864.8--- is not feasible. Nevertheless, it is significant to 
confirm a second DIB close to the strong one $\lambda$862 in such a big spectroscopic survey, which enables us to use them as 
tracers of the ISM environment if they have different origins, or facilitates the study of their origin if they share a common
carrier (with two DIBs, their relative strength and central wavelengths could provide clues as to their carrier). Dedicated
research into these two DIBs will be carried out using all available RVS spectra and the results will be published in next {\it Gaia}
data release.

\section{Conclusions} \label{sect:conclusion}

Based on the measurements in 1103 stacked {\it Gaia}--RVS spectra, we provide solid confirmation of the DIB around 864.8\,nm through 
its correlation with $\lambda$862 and its clear and broad profile in the stacked spectra. $\lambda$864.8 is a very broad and shallow 
DIB. The FWHM of $\lambda$864.8 is estimated to be $1.62\,{\pm}\,0.33$\,nm. Using 43 high-quality measurements toward GC 
($|\ell|\,{\leqslant}\,10^{\circ}$, $|b|\,{\leqslant}\,10^{\circ}$), the rest-frame wavelength of $\lambda$864.8 is determined as 
$\lambda_0\,{=}\,864.53\,{\pm}\,0.14$\,nm in the vacuum, which is smaller than previous reported measurements. $\rm EW_{862}$ correlates 
better with $\rm E(BP-RP)$ than $\rm EW_{864.8}$. 

Our work shows the power of using a large set of cool-star spectra to study the DIBs in the ISM. The extremely small depth of these 
lines ($\rm CD_{864.8}\,{\lesssim}\,3$\%) and the ability to assess their properties is a clear demonstration of the quality of {\it 
Gaia}--RVS spectra as published in Gaia DR3.

\begin{acknowledgements}
  This work has made use of data from the European Space Agency (ESA) mission {\it Gaia} (\url{https://www.cosmos.esa.int/gaia}), 
  processed by the {\it Gaia} Data Processing and Analysis Consortium (DPAC, \url{https://www.cosmos.esa.int/web/gaia/dpac/consortium}). 
  Funding for the DPAC has been provided by national institutions, in particular the institutions participating in the {\it Gaia} 
  Multilateral Agreement. This work was supported in part by: the German Aerospace Agency (DLR) grant 50 QG 2102; SFB 881 ``The Milky 
  Way System'' of the German Research Foundation (DFG).
\end{acknowledgements}

\bibliographystyle{aa}
\bibliography{reference.bib}

\appendix

\section{ADQL queries} \label{queries}
\noindent
\begin{verbatim}
SELECT *
FROM gaiadr3.astrophysical_parameters AS gaia 
INNER JOIN gaiadr3.gaia_source AS xmatch
ON gaia.source_id = xmatch.source_id 
WHERE (xmatch.has_rvs='T' AND gaia.teff_gspspec > 3500 AND gaia.teff_gspspec < 7500 
AND gaia.logg_gspspec < 6 AND xmatch.parallax > 0.0 AND 1/xmatch.parallax < 6  AND
xmatch.radial_velocity_error < 5.0 AND  xmatch.rv_expected_sig_to_noise > 20.0 AND
gaia.mh_gspspec is NOT NULL)
\end{verbatim}


\clearpage

\end{CJK*}

\end{document}